\documentclass[12pt]{article}

\usepackage[utf8]{inputenc}
\usepackage{charter}
\usepackage{fullpage}
\usepackage{hyperref}
\usepackage{graphicx}
\usepackage{lipsum}
\usepackage[sort&compress,numbers]{natbib}
\usepackage{hyperref}
\hypersetup{hidelinks}
\usepackage[total={6.5in,9in},top=1in,headsep=0.1in,headheight=1in]{geometry}
\usepackage{enumitem,amssymb}
\usepackage[usenames,dvipsnames]{xcolor}
\usepackage{xfrac}
\usepackage{bm}
\usepackage{array}
\usepackage{comment}

\newcommand\snowmass{
\begin{center}
  \rule[-0.2in]{\hsize}{0.01in}\\
  \rule{\hsize}{0.01in}\\
  \vskip 0.1in
  Submitted to the Proceedings of the US Community Study\\ 
  on the Future of Particle Physics (Snowmass 2021)\\
  \rule{\hsize}{0.01in}\\
  \rule[+0.2in]{\hsize}{0.01in}\\[-2em]
\end{center}
}

\usepackage[firstpage=true]{background}
\backgroundsetup{contents={\parbox{6.5in}{\snowmass}}, scale=1,placement=top,opacity=1,color=black,position={3.25in,1.2in}}

\usepackage{fancyhdr}
\fancypagestyle{plain}{%
  \fancyhf{}%
  \fancyhead[C]{}
  \fancyfoot[C]{\thepage}
}

\fancypagestyle{empty}{%
  \fancyhf{}%
  \fancyhead[C]{{\it Cosmology and fundamental physics from large-scale structure}}
  \fancyfoot[C]{\thepage}
}
\newcommand{\MW}[1]{\textcolor{magenta}{[MW: #1]}}

\title{Snowmass2021 Cosmic Frontier White Paper: Cosmology and Fundamental Physics from the three-dimensional Large Scale Structure}
\date{\today}

\usepackage{authblk}

\author[1,2]{Simone Ferraro}
\author[2,1]{Noah Sailer}
\author[3]{An\v{z}e Slosar}
\author[2,1]{Martin White}
\author[\space]{\\ for the Snowmass 2021 Cosmic Frontier 4 Topical Group}

\affil[1]{Lawrence Berkeley National Laboratory, One Cyclotron Road, Berkeley, CA 94720, USA}
\affil[2]{Berkeley Center for Cosmological Physics, Department of Physics,
University of California, Berkeley, CA 94720, USA}
\affil[3]{Physics Department, Brookhaven National Laboratory, Upton, NY 11973, USA}

\begin{document}

\maketitle

\begin{abstract}
Advances in experimental techniques make it possible to map the high redshift Universe in three dimensions at high fidelity in the near future. This will increase the observed volume by many-fold, while providing unprecedented access to very large scales, which hold key information about primordial physics. Recently developed theoretical techniques, together with the smaller size of non-linearities at high redshift, allow the reconstruction of an order of magnitude more ``primordial modes'', and should improve our understanding of the early Universe through measurements of primordial non-Gaussianity and features in the primordial power spectrum.  In addition to probing the first epoch of accelerated expansion, such measurements can probe the Dark Energy density in the dark matter domination era, tightly constraining broad classes of dynamical Dark Energy models. The shape of the matter power spectrum itself has the potential to detect sub-percent fractional amounts of Early Dark Energy to $z \sim 10^5$, probing Dark Energy all the way to when the Universe was only a few years old. The precision of these measurements, combined with CMB observations, also has the promise of greatly improving our constraints on the effective number of relativistic species, the masses of neutrinos, the amount of spatial curvature and the gravitational slip.  Studies of linear or quasi-linear large-scale structure with redshift surveys and the CMB currently provide our tightest constraints on cosmology and fundamental physics.  Pushing the redshift and volume frontier will provide guaranteed, significant improvements in the state-of-the-art in a manner that is easy to forecast and optimize. 
\end{abstract}

\section{Introduction}

The inhomogeneous Universe, as probed by fluctuations in the cosmic microwave background (CMB) radiation or surveys of large-scale structure (LSS), provides one of our best windows on fundamental physics at ultra-high energies.  The tightest constraints on dark energy, mass limits on light dark matter particles, models of inflation, neutrino masses and light relic particles all come from one or both of these measurements.  Continuous advances in detector technology and experimental techniques are pushing us into a new regime, enabling mapping of large-scale structure in the redshift window $2<z<6$ using both relativistic and non-relativistic tracers.  This will allow us to probe the metric, particle content and \emph{both} epochs of accelerated expansion (Inflation and Dark Energy domination) with high precision in a regime that is not theory limited.

Cosmological constraints from the CMB and LSS are well developed and can be accurately forecast for modes which are in the linear or quasi-linear regimes.  They include constraints on the expansion history and curvature \cite{Slosar19c}, primordial non-Gaussianity \cite{Meerburg19, Snowmass2021:Inflation}, features in the power spectrum (primordial \cite{Slosar19b, Snowmass2021:Inflation} or induced \cite{Hill20,Chen20b}) or running of the spectral index \cite{Ferraro19,Castorina20}, dark energy in the approach to matter domination \cite{Slosar19a}, dark matter interactions \cite{CMBS4,Gluscevic19}, light relics and neutrino mass \cite{Dvorkin19,Green19} and modified gravity \cite{Jain10,Joyce15}.  The theoretical community has highlighted many ways in which we could observe evidence for beyond standard model (BSM) physics, however at present none of these avenues appears more compelling than the others.  While we do not know from theory which of these many probes is most likely to turn up evidence of BSM physics, well understood phenomenology allows us to forecast where our sensitivities will be highest and our inference cleanest.  This amounts to maximizing our $S/N$ for new physics not by trying to find the location where $S$ is maximized but rather identifying where $N$ is minimized.  To work where the inference is cleanest and the noise lowest we should push into the new frontier at high redshift.

Moving to higher redshift allows us to take advantage of four simultaneous trends.  (1) A wider lever arm in redshift leads to rotated degeneracy directions, tightening constraints.  (2) The volume on the past lightcone increases dramatically, leading to much tighter constraints on sample-variance limited modes and a longer lever arm in scale. (3) The degree of non-linearity is smaller, and the field is better correlated with the early Universe and less affected by astrophysical processes.  (4) Very high precision perturbative models built around principles familiar from high-energy particle physics become increasingly applicable.  Indeed at high redshift, with large volume surveys, we increasingly probe long wavelength modes which are linear or quasi-linear and thus carry phase information from the early Universe, before it has been permanently lost to non-linear evolution.

Quantifying the information on primordial physics that could come from future large-$N_{\rm mode}$ surveys is the key goal of this white paper.  Since non-linear evolution, astrophysical processes and noise cause decorrelation with the initial conditions, and hence loss of information on small scales,  the effective number of modes that are correlated with the initial conditions increases with redshift.  We quantify this using a ``Primordial Physics Figure of Merit'' (Primordial FoM), introduced in Appendix \ref{sec:FOM}, and which is proportional to the number of modes that can be measured and are correlated with the initial conditions, properly accounting for noise in the measurement and decorrelation due to non-linearities.

After briefly introducing the probes and facilities capable of implementing the large-N program in \S\ref{sec:probes}, we lay out our forecasting assumptions in \S\ref{sec:forecasting}.  The rest of the white paper explores the consequences for constraining early-Universe physics from the billions of linear or quasi-linear modes that are available (in principle) to a high redshift large-scale structure survey.  Our ability to constrain the density fluctuations is discussed in \S\ref{sec:Pofk}.
We discuss constraints on inflation in \S\ref{sec:primordial} and on the expansion history in \S\ref{sec:expansion}.  In \S\ref{sec:DE} we show how observing the high-$z$ universe can be transformative in our study of Dark Energy.  Constraints on neutrino masses, light relics and other extensions of our standard model as discussed in \S\ref{sec:extensions}.
The synergies of high-redshift galaxy surveys and future wide-area and low noise CMB surveys (such as CMB-S4 \cite{CMBS4}, or more ambitious proposals like CMB-HD \cite{Sehgal:2019ewc} or PICO \cite{NASAPICO:2019thw}) are explored in \S\ref{sec:synergies}.

\section{Cosmological probes at high redshift}
\label{sec:probes}

\subsection{Opportunities}

We anticipate that we will first map large-scale structure at high redshift with deep imaging surveys and CMB lensing, leveraging the community's investments in Rubin LSST \cite{LSST}, CMB-S4 \cite{CMBS4} and other facilities.  Upcoming experiments will already be able to provide percent-level constraints on the growth of structure above $z\approx 2$ using cross-correlations \cite{Wilson:2019brt}.  Limited spectroscopic follow-up would enable constraints on $dN/dz$ for interpreting the cross-correlations.  The eventual goal, however, would be to unlock the full power of a 3D survey using spectroscopic redshifts.  While it is not our goal to advocate for a particular observational approach, we will mention that there are three techniques for tracing large-scale structure for which we can make spectroscopic observations with next-generation facilities.  First, building upon deep imaging from LSST \cite{LSST} and Euclid \cite{EUCLID18}, we could target\footnote{To what extent follow-on imaging by LSST and Euclid ``extensions'' would enhance this science case remains an open question.} Lyman Break Galaxies \cite{Wilson:2019brt}.  These galaxies are abundant, well studied and well understood, representing massive, actively star-forming galaxies with a luminosity approximately proportional to their stellar mass.  A sizeable fraction have bright emission lines which could facilitate obtaining a redshift \cite{Wilson:2019brt}.  Early star-forming galaxies can also be dusty, absorbing much of the stellar optical light and re-emitting in the IR. Rest-frame far-IR lines such as the CO rotational transitions and the [C{\sc ii}] fine-structure line can similarly be used for redshift determination using spectroscopy at millimeter wavelengths \cite{Kovetz17, Karkare2021:LIM}. These lines are bright and have been detected in individual galaxies out to $z\sim 7$, but the possibility of line confusion is a concern.  Finally, the majority of the baryonic matter in the Universe is in the form of Hydrogen.  Neutral hydrogen, predominantly in galaxies, emits through the $21\,$cm hyperfine, magnetic dipole, transition (see also a dedicated Snowmass 2021 White Paper \cite{CF5_21cm}).  This line is weak and at long wavelength, requiring large collecting area to detect it, but there is little absorption or confusion if the line can be detected \cite{Slosar19a,Castorina20}.

\subsection{Large-scale structure experiments}
\label{sec:experiments}

Future experiments will build upon the strong legacy of spectroscopy from DESI \cite{DESI}, Euclid \cite{EUCLID18} and SPHEREx \cite{Dore14} that will measure redshifts of line emitting galaxies over at least 10,000 square degrees out to $z\simeq 2.7$.  The combination of these experiments, probing the optical and near-infrared, will measure most of the linear modes accessible out to redshifts between 1 and 2 with near sample variance dominated precision.

Future galaxy redshift surveys can extend the redshift frontier significantly.  While DESI-II \cite{DESI2} would begin this program, future facilities, including MegaMapper \cite{MegaMapper}, the Maunakea Spectroscopic Explorer (MSE; \cite{MSE}) and SpecTel \cite{SpecTel}, would be required to truly exploit it. A spectroscopic road map to achieve the science goals outlined in this paper has been recently proposed \cite{DESI:2022lza}. MegaMapper is a proposed highly-multiplexed spectroscopic instrument designed to cover $2<z<5$. It combines a $6.5\,$m telescope aperture with $\sim20$K fibers.  MegaMapper's location at the Las Campanas Observatory will ensure full overlap with the Rubin Observatory, covering roughly 14,000 square degrees. MSE has a 11.25 m mirror and a 1.5 square degree field of view, with the capability of observing thousands of astronomical objects simultaneously using up to $\sim 16,000$ fibers (depending on the design). MSE will measure spectra from 360 to $950\,$nm, detecting galaxies out to $z=3$ over 10,000 square degrees. SpecTel is a proposed spectroscopic survey in the southern hemisphere that would couple an $11.4\,$m dish (with a 5 square degree FoV) to 15,000 fibers. It would take spectra in the range 360 to $1330\,$nm.

Millimeter-wave line intensity mapping (LIM) from $80-300\,$GHz is capable of probing $0<z<10$ using multiple CO lines and [C{\sc ii}] (see the dedicated Snowmass white paper \cite{Karkare2021:LIM}).  While pathfinder instruments are still being fielded, strong heritage in scaling up detector counts and observing strategy from the CMB community will allow significantly more powerful surveys to be deployed on a 10-15 year timescale.  Moreover, existing $5-10\,$m-class CMB facilities at excellent mm-wave observing sites (South Pole, Atacama desert) capable of surveying $\sim 70\%$ of the sky could be repurposed, allowing for deployment at modest cost.  For example, a fully-populated receiver filling the field of view of the South Pole Telescope could field 2800 densely-packed $R\sim 300$ on-chip spectrometers, while a CMB-S4 Large-Aperture Telescope could host 34000.

The Packed Ultrawideband Mapping Array (PUMA) is a $21$-cm interferometer capable of mapping large-scale structure out to $z\approx 6$ \cite{Slosar19a,Castorina20}.  The full instrument (PUMA-32K) consists of an array of 32,000 six-meter parabolic dishes, steerable in declination, with 0.2 - 1.1GHz bandwidth receivers located on the sites of a hexagonal close-packed lattice with 50\% occupancy.  A smaller, 5,000-dish version of the experiment (PUMA-5K) with reduced scientific reach is also envisioned.  In its full configuration, and with a five year survey, PUMA's total noise will be equivalent to the sampling (Poisson) noise from a spectroscopic survey of 2.9 billion galaxies out to $z=6$.

In what follows, we will make explicit forecasts for galaxy surveys and 21\,cm intensity mapping to compare the strengths of different techniques.  Since the forecasts for those and for LIM are not performed using the same forecasting machinery \cite{Moradinezhad19,Moradinezhad21}, we do not report the LIM forecasts directly.  We note that LIM will share many of the same features with 21\,cm that stem from the need to perform foreground filtering. In detail, however, it has different analysis issues: while 21\,cm is most likely limited by our ability to calibrate the instrument, the microwave LIM main challenges are interloper lines and limited spectroscopic resolution.

We will use the number density, bias assumptions and noise levels from Section 2 of \cite{Sailer:2021}. While these are our current best estimates, the properties of galaxy populations at high redshift are subject to considerable uncertainty, and our understanding should improve considerably in the near future thanks to observations from DESI, HSC, HETDEX and other surveys.  Our uncertainties on the manner in which H{\sc i} traces the cosmic web at high redshift are even larger, so constraints from pathfinder experiments will be even more important.

\begin{table}
\footnotesize
\newcolumntype{x}{>{\centering\arraybackslash\hspace{0pt}}p{1.2cm}}
\newcolumntype{y}{>{\centering\arraybackslash\hspace{0pt}}p{2.0cm}}
\newcolumntype{z}{>{\centering\arraybackslash\hspace{0pt}}p{2.7cm}}
\hspace*{-1.8cm}  \begin{tabular}{|p{2.2cm}|x|z|y|y|x|y|z|}
                  \hline &&&&&&&
                  \\
                  & Experiment type & Concept & Redshift Range & Primordial FoM & Time-scale & Technical Maturity & Comments \\
\hline
DESI & \scriptsize spectro & \scriptsize 5000 robotic fiber fed spectrograph on 4m Mayall telescope  & \scriptsize $0.7<z<2.0$ & 0.88 & now & operating & \\
\hline
Rubin LSST & \scriptsize photo &  \scriptsize \textit{ugrizy} wide FoV imaging on a 6.5m effective diameter dedicated telescope & \scriptsize $0<z<3$ & - & 2025-2035 & on schedule & \scriptsize Targeting survey for next generation spectroscopic instruments \\
\hline
SPHEREx & \scriptsize narrowband & \scriptsize Variable Linear Filter imaging on 0.25m aperture from space & \scriptsize $0<z<4$ & - & 2024 & on schedule & \scriptsize Focus on primordial non-Gaussianity \\
\hline
\hline
MSE+${}^\dagger$ & \scriptsize spectro & \scriptsize up to 16,000 robotic fiber fed spectrograph on 11.25\,m telescope &\scriptsize  $1.6 < z < 4$ (ELG+LBG samples) & $< 6.1$ & 2029- & high & \\
\hline
MegaMapper & \scriptsize spectro & \scriptsize 20,000 robotic fiber fed spectrograph on 6m Magellan clone & \scriptsize $2<z<5$ & 9.4 &2029- & high &  \scriptsize Builds upon existing hardware and know-how \\
\hline
SpecTel${}^\dagger$ & \scriptsize spectro & \scriptsize 20,000-60,000 robotic fiber fed spectrograph on a dedicated 10m+ class telescope & \scriptsize $1<z<6$ & $<23$ & 2035- & medium  &  \scriptsize Potentially very versatile next generation survey instruments\\
\hline
PUMA & \scriptsize 21\,cm & \scriptsize 5000-32000 dish array focused on intensity 21\,cm intensity mapping &  \scriptsize $0.3 <z<6$ & 85 / 26 (32K / 5K optimistic) & 2035- & to be demonstrated  & \scriptsize Very high effective number density, but $k_\parallel$ modes lost to foregrounds \\
\hline
mm-wave LIM concept & \scriptsize microwave LIM & \scriptsize 500-30000 on-chip spectrometers on existing 5-10m telescopes, 80-300\,GHz with R$\sim$300-1000 & \scriptsize $0<z<10$ & up to 170 &  2035 - & to be demonstrated & \scriptsize CMB heritage, can deploy on existing telescopes, signal uncertain, $k_\parallel$ modes lost to foregrounds \& resolution \\
\hline
  \end{tabular}
  \caption{Table comparing current and next generation experiments capable of performing 3D mapping of the Universe. The upper part of the table shows existing and funded experiments, while the lower part is focused on proposed future facilities. See \cite{Sailer:2021} for further details. ${}^\dagger$ We have computed the FoM for MSE and SpecTel assuming they performed a full time LBG/LAE survey -- such a survey was not part of their proposals and those collaborations have not committed to doing any such survey.  For their proposed surveys the FoM is significantly lower. }
  \label{tab:experiments}
\end{table}

\section{Forecasting assumptions}
\label{sec:forecasting}

We follow the forecasting formalism of \cite{Sailer:2021} to obtain the cosmological forecasts, using the public code \verb|FishLSS|\footnote{\url{https://github.com/NoahSailer/FishLSS}}.
We denote the nonlinear redshift-space power spectrum of the matter tracer by $P_{gg,i}(z) = P_{gg}(k_i,\mu_i,z)$. We divide the experiment into redshift bins $\{B_\alpha;\alpha=1,2,\cdots\}$, then calculate the Fisher matrix for each bin, given a set of parameters $\{p_a\}$:
\begin{equation}
    F_{ab}(B_\alpha)
    =
    \sum_{ij}
    \frac{\partial P_{gg,i}(z)}{\partial p_a}
    C^{-1}_{ij}(z)
    \frac{\partial P_{gg,j}(z)}{\partial p_b}
    \bigg|_{z=z_\alpha}
    \\
    \textrm{ where }
    \quad
    C_{ij}(z)
    = 
    \delta^K_{ij}
    \frac{4\pi^2}{k_i^2 V_\alpha dk_i d\mu_i}
    \left[
    P_{gg,i}(z)
    +
    N_i(z)
    \right]^2.
\label{eq:partial_fisher}
\end{equation}
Where $V_\alpha$ is the comoving volume of bin $B_\alpha$, at redshift $z_\alpha$. $N_i(z)$ is the noise, which is assumed to be a constant within bin $B_\alpha$.  

The final Fisher matrix is obtained by summing over redshift bins:
\begin{equation}
    F_{ab} =  \sum_\alpha F_{ab}(B_\alpha).
\end{equation}
We assume that the bins are large enough to make the covariances between different redshift bins negligible. 
For all experiments we assume $k_\textrm{min} = \textrm{max}\big( 0.003\,h\,\textrm{Mpc}^{-1}, \pi/V^{1/3}_\alpha \big)$.  Unless explicitly stated otherwise, for all experiments we assume $k_\textrm{max}(z) = k_{\rm nl}(z)$, where $k_{\rm nl}(z) = 1/\Sigma(z)$ is the non-linear scale set by the RMS displacement in the Zel'dovich approximation.
When forecasting H{\sc i} surveys, we also include a foreground-wedge \cite{CVDE-21cm} that further constrains the limits of integration:
\begin{equation}
    k_\parallel 
    >
    \textrm{max}\left[
    k_\perp
    \frac{\chi(z)H(z)}{c(1+z)} \sin(\theta_w(z))
    \,\,,\,\,k_\parallel^\textrm{min}\right]
    \quad
    \textrm{where}
    \quad 
    \theta_w(z) =
    N_w
    \frac{1.22}{2 \sqrt{0.7} } 
    \frac{\lambda_{\rm obs}}{D_\textrm{phys}}.
\end{equation}
$\lambda_{\rm obs}=(1+z)\,21\,$cm is the observed wavelength and $D_\textrm{phys} = 6\,$m is the physical diameter of the dish. For each H{\sc i} survey we consider ``optimistic'' and ``pessimistic'' foreground cases, in which $N_w=1$ , $k_\parallel^\textrm{min}=0.01\,h\,{\rm Mpc}^{-1}$ and $N_w=3$ , $k_\parallel^\textrm{min}=0.1\,h\,{\rm Mpc}^{-1}$ respectively. This is consistent with the definitions in refs.~\cite{CVDE-21cm,Sailer:2021}.

The linear power spectrum is computed with the code \verb|CLASS|\footnote{\url{https://github.com/lesgourg/class_public}}, while non-linear corrections are obtained with the \verb|velocileptors|\footnote{\url{https://github.com/sfschen/velocileptors}} implementation of one-loop Lagrangian Perturbation Theory (LPT) \cite{Chen19} including effective field theory terms \cite{Carrasco:2012cv, BNSZ12, Sen14, Porto:2013qua, VlaWhiAvi15}.
In particular, we consider a model with linear, quadratic and shear bias, together with 3 counterterms $\alpha_{0, 2, 4}$ and 3 stochastic contributions $N_{0,2,4}$ which are marginalized over (see \cite{Sailer:2021} for details). For 21cm, we further marginalize over $\Omega_{\rm HI}(z)\equiv \rho_{\rm HI}(z)/\rho_c(z)$, the cosmic density of neutral hydrogen.

\section{Power Spectrum Measurements}
\label{sec:Pofk}

\begin{figure}[!h]
    \centering
    \includegraphics[width=0.8\linewidth]{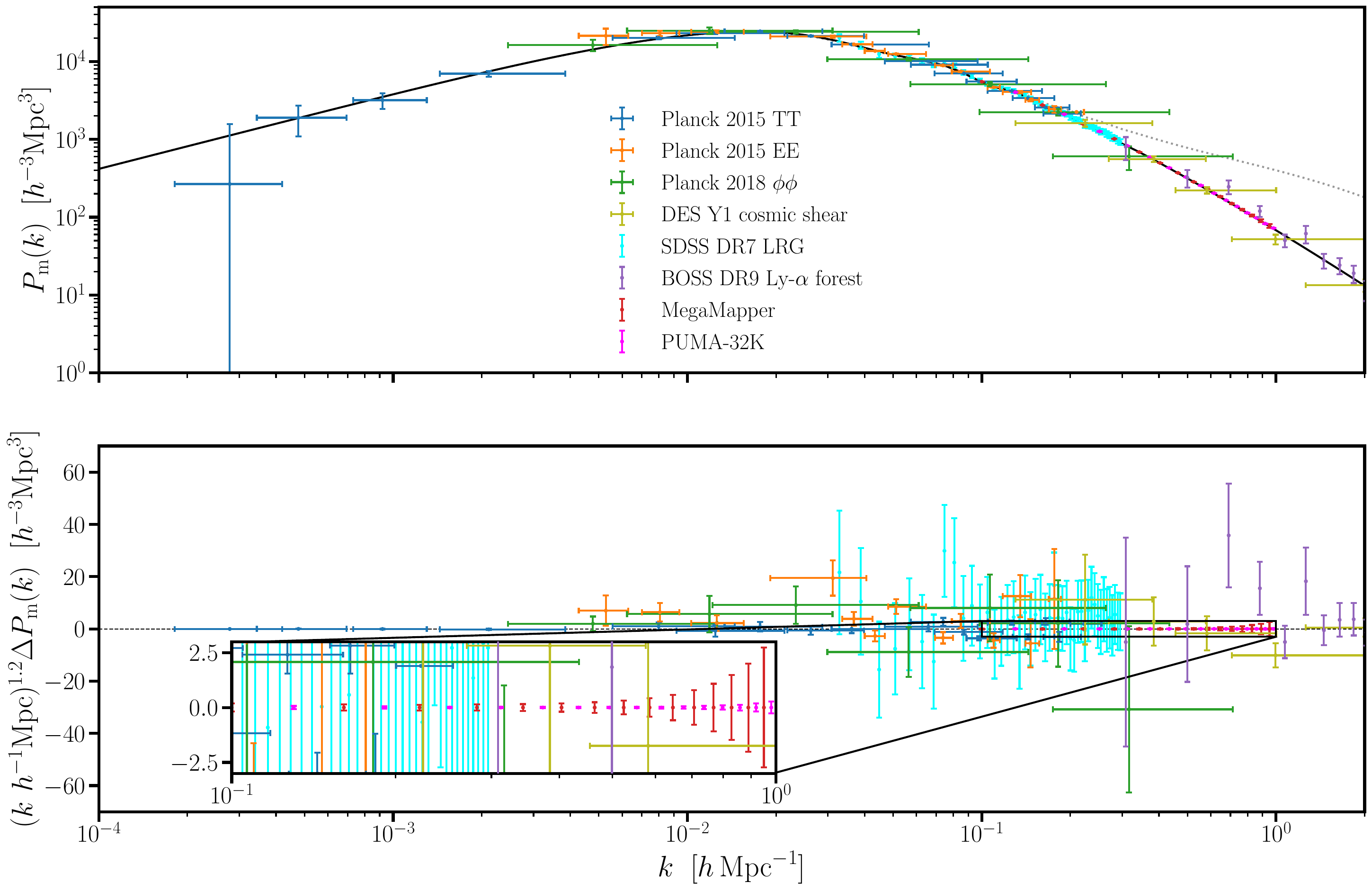}
    \caption{Measurements of the linear matter power spectrum at $z=0$. For both MegaMapper and PUMA-32K we show constraints for 15 linearly space $k$-bins between $0.1\,\,h\,\text{Mpc}^{-1}\lesssim k\lesssim 1\,\,h\,\text{Mpc}^{-1}$. This figure is adapted from refs. \cite{PlanckLegacy18, Chabanier_2019, Sailer:2021}.  } 
\label{fig:marius_plot}
\end{figure}

Thanks to the large volume available at high redshift and the availability of a sufficient number density of tracers, the matter power spectrum will be measured with exquisite precision by the next generation of cosmological surveys. Figure \ref{fig:marius_plot} shows current and future constraints on the three-dimensional power spectrum, extrapolated to $z = 0$ using linear theory. This measurement, when performed as a function of redshift, allows us to constrain a large number of effects in both the high and low redshift Universe as we'll discuss in the next few sections. 

We note that excellent constraints on the projected power spectrum, integrated along the line of sight, can be obtained by future galaxy lensing (for example Rubin Observatory \cite{LSST}) or wide-field CMB lensing (for example CMB-S4 \cite{CMBS4}, CMB-HD \cite{Sehgal:2019ewc} or PICO \cite{NASAPICO:2019thw}). Those measurements have the advantage of potentially being easier to model to slightly smaller scales (due to their insensitivity to non-linear bias effects and lower noise), and are excellent at measuring the amplitude of the small scale power. On the other hand, the projection limits the number of modes that can be measured (and hence the precision) and will wash out detailed features in the power spectrum, making it harder to extract  information about the expansion or the imprints of inflation. Therefore, measurement of both the projected and three-dimensional power spectra are highly complementary. In addition, the cross-correlation between the two unlocks the synergies explored in \S \ref{sec:synergies}, allowing for greater statistical power and reduced systematics, further strengthening the case for pursuing both of them in parallel.

\begin{figure}[!h]
    \centering
    \includegraphics[width=0.8\linewidth]{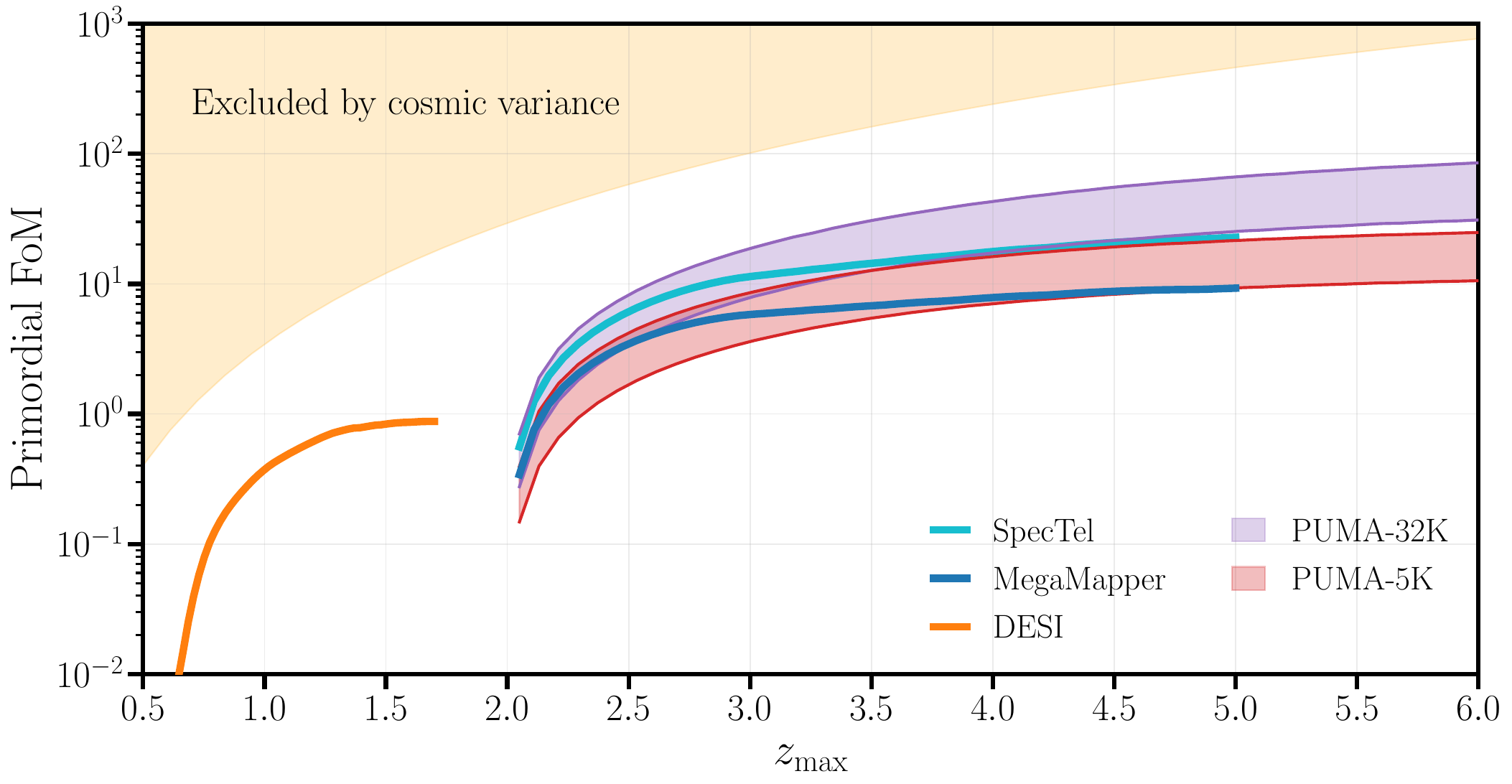}
    \caption{Primordial FoM $\equiv 10^{-6}\,N_\text{modes}$ as a function of $z_\text{max}$ for DESI, PUMA (-5K and -32K), MegaMapper and SpecTel.  For DESI we include only the ELGs. For PUMA, we consider both optimistic and pessimistic foreground models, which are the boundaries of the shaded regions. The boundary of the shaded orange region is the cosmic variance limit for an all-sky survey, assuming $b(z)=1$.
    } 
\label{fig:figure_of_merit}
\end{figure}

The uncertainty on the power spectrum in a bin of width $\Delta k$ is given by $P_t(k)\sqrt{2/N(k)\Delta k}$, where $N(k)$ is the number of modes between $k$ and $k+\Delta k$ and $P_t(k)$ is the total power (including any noise contribution). However, not all modes are equally useful in constraining primordial physics as we discuss next.

\section{Primordial Physics}
\label{sec:primordial}

Since the photon free path is microscopic before the decoupling of light and matter at recombination, the Cosmic Microwave Background is the earliest epoch of universe we can observe directly. Everything else must be inferred indirectly. In this paper we take primordial physics to mean any information about the state of the early universe that is encoded in the statistics of fluctuations of the late universe. For a review of primordial physics that can be probed by LSS experiments, as well as the theoretical motivations, we refer the reader to the Snowmass paper on Inflation \cite{Snowmass2021:Inflation}.

The large number of primordial modes that can be reconstructed by observing the LSS at $z>2$ will allow future LSS experiments to outperform CMB measurements in constraining primordial physics. As shown in Fig.\ \ref{fig:marius_plot}, the (linearly-evolved) matter power spectrum can be measured to unprecedented precision, and in turn will reveal any deviation from a nearly scale-invariant primordial power spectrum, predicted by the simplest models of single-field slow-roll inflation. For example, ref.\  \cite{Sailer:2021} shows that the amplitude of a sinusoidal modulation in the power spectrum (generically expected from step-like features in the inflationary potential), can be constrained by MegaMapper or PUMA-32K up to one order of magnitude better than near-term spectroscopic surveys.

\begin{figure}[!h]
    \centering
    \includegraphics[width=0.6\linewidth]{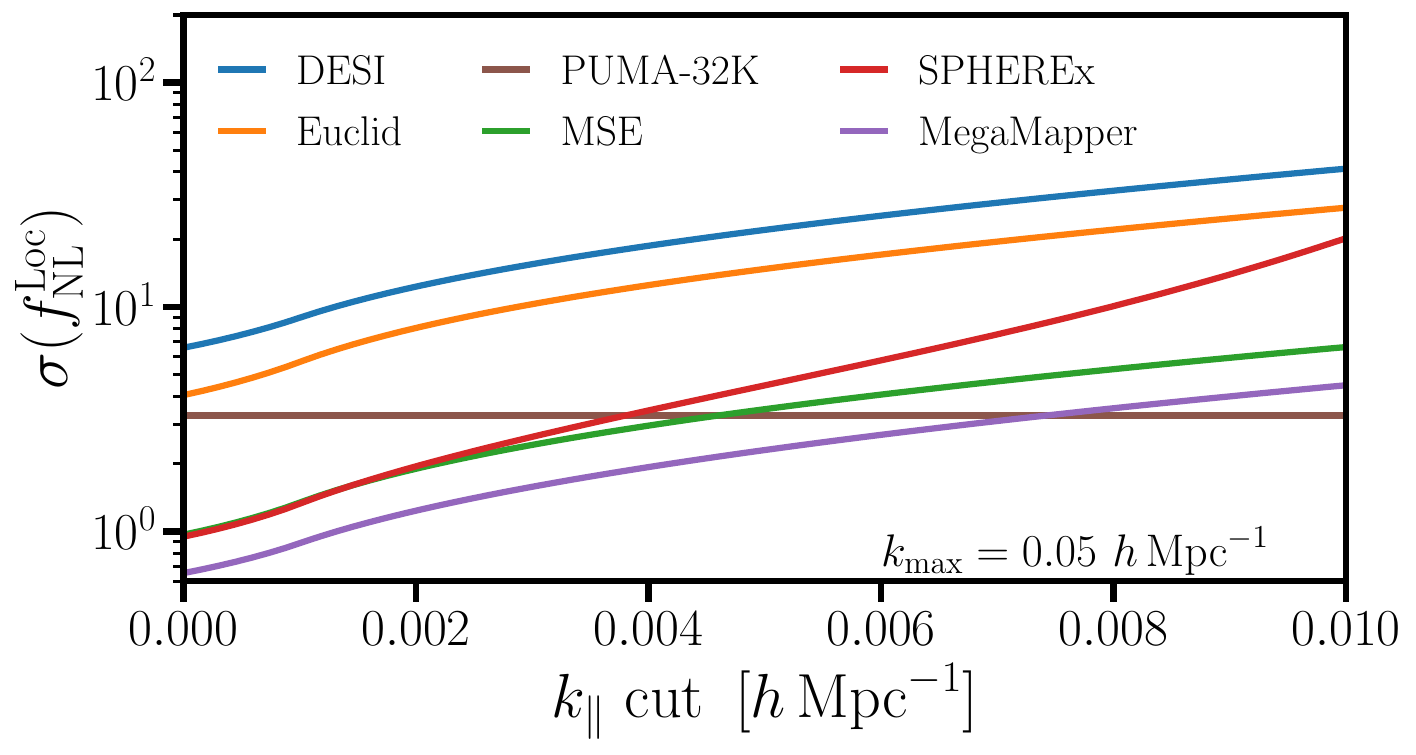}
    \caption{Constraints on the amplitude of local primordial non-Gaussianity with a cut on $k_\parallel$ modes, to control foregrounds and angular systematics. For SPHEREx we assume a redshift error of $\sigma_z/(1+z)=0.05$, while for all other surveys we set $\sigma_z=0$.  DESI refers to the emission line galaxy (ELG) sample from the DESI survey. The PUMA-32K constraint (horizontal line) assumes an optimistic foreground scenario, which already includes a $k_\parallel > 0.01\,\,h\,\text{Mpc}^{-1}$ cut. For all surveys we set $k_\text{max} = 0.05\,\,h\,\text{Mpc}^{-1}$ and $k_\text{min}=0.001\,\,h\,\text{Mpc}^{-1}$. 
    }
\label{fig:fnl_vs_kparmin}
\end{figure}

Similarly, primordial non-Gaussianity of the local type can be measured by looking for scale-dependent bias on very large scales, and hence constrained from the power spectrum alone \cite{Dalal:2007cu, Ferraro:2014jba}. Fig.\ \ref{fig:fnl_vs_kparmin} shows that high-$z$ surveys can cross the important theoretical threshold $\sigma(f_{\rm NL}^{\rm loc}) \lesssim 1$ (generically separating single-field and multi-field inflationary models \cite{Snowmass2021:Inflation}), reaching beyond what's achievable with the CMB, which is intrinsically 2-dimensional and limited by the cosmic variance.

If large-scale systematics can be controlled, MegaMapper can achieve\footnote{It has been pointed out \cite{Barreira:2022sey} that scale-dependent bias in LSS surveys provides a direct measurement of $b_\phi f_{NL}^{\rm local}$, where $b_\phi$ is the response of the galaxy density to changes in local $f_{\rm NL}$. In this work we fix $b_\phi$ to its fiducial peak-background split value $b_\phi \approx 2\delta_c (b -1)$, where $\delta_c$ is the critical density, and $b$ is the (linear) galaxy bias \cite{Dalal:2007cu}. However, forecasts can easily be rescaled for any chosen value of $b_\phi$.} $\sigma(f_{\rm NL}^{\rm loc}) \approx 0.7$ from the power spectrum alone, with potential improvements coming from measuring the bispectrum as well \cite{Ferraro:2019uce, Karagiannis:2018jdt}. PUMA is loosing the largest-scale modes due to foreground filtering, but should nevertheless be competitive also for local non-Gaussianity through bispectrum constraints \cite{2020JCAP...11..052K}. We note that a combination of MegaMapper for large-scale modes and PUMA for small-scale modes over the same volume should be particularly constraining bispectrum combination that should additionally be more robust to systematics.

For other ``shapes'' of non-Gaussianity such as equilateral or orthogonal, the effects on the power spectrum are largely degenerate with the bias parameters and a measurement of the bispectrum is required. Both PUMA and MegaMapper, should be able to improve the current bounds on Equilateral and Orthogonal non-Gaussianity by a factor $\approx 2 - 3$ from measurements of the LSS bispectrum \cite{Ferraro:2019uce, Slosar19a}.

\section{Expansion Constraints}

\begin{figure}[!h]
\centering
\includegraphics[width=\linewidth]{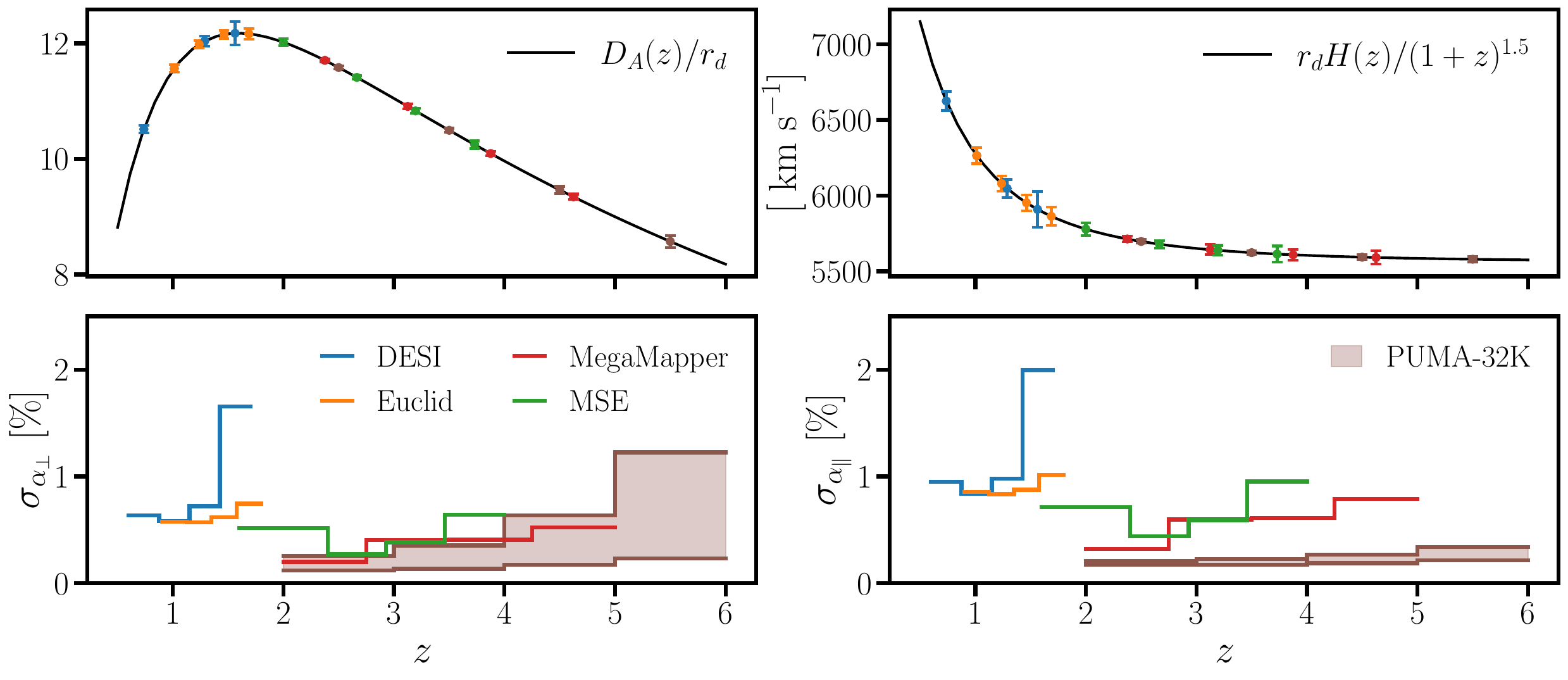}
\caption{
Error on the parameters $\alpha_\perp, \alpha_\parallel$ from the reconstructed power spectrum, which can be interpreted as relative errors on $D_A(z)/r_d$ and $r_d H(z)$ respectively.  The line for DESI is specifically for the ELG sample. The boundaries of the shaded regions denote optimistic/pessimistic foreground assumptions for the 21-cm surveys. In the top panels we show the errorbars for the optimistic case.
}
\label{fig:distance_constraints}
\end{figure}

The fundamental degrees of freedom of the Friedmann metric, geometry and expansion history, can be tightly constrained by measures of the distance-redshift relation(s).
Measurements of Baryon Acoustic Oscillations (BAO) constrain the expansion parameters $D_A(z)/r_d$ and $H(z)r_d$ where $D_A(z)$ is the angular diameter distance to redshift $z$, $H(z)$ is the Hubble parameter and $r_d$ is the sound horizon at the drag epoch.
These in turn can be used to constrain the expansion history and rate, informing us on the composition of the Universe and time-dependence of the various energy sources, as well as curvature.
Figure \ref{fig:distance_constraints} shows the fractional error on $D_A(z)/r_d$ and $H(z)r_d$ (as the error on the Alcock-Paczynski parameters $\alpha_\perp$ and $\alpha_\parallel$ respectively). Remarkably, future LSS experiments will allow the measurement of expansion parameters to sub-percent precision all the way to $z \gtrsim 5$, thus fully spanning the Dark Energy dominated era and extending well into matter domination. As we explain in the next section, this is precisely what's needed to constrain large classes of dynamical Dark Energy models, and can also shed light on other energy components such as curvature or modifications to General Relativity. Notably the constraints on curvature alone will improve by more than a factor of two with future high-redshift surveys (see Fig.\  \ref{fig:extensions_to_base_model}).

\label{sec:expansion}
\ \\

\section{(Early) Dark Energy}
\label{sec:DE}

\begin{figure}[!h]
\centering
\includegraphics[width=\linewidth]{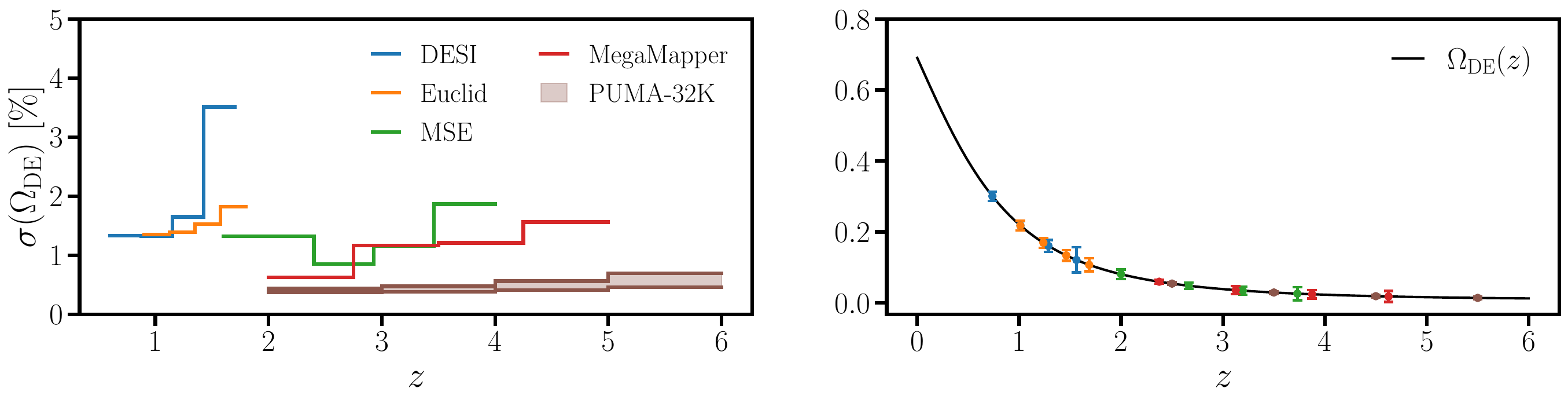}
\caption{The absolute error on the dark energy fraction ($\Omega_\text{DE}$) achievable by current and future LSS surveys, thanks to direct expansion measurements. From \cite{Sailer:2021}. }
\label{fig:Omega_DE}
\end{figure}

Observing the high-$z$ universe through measurements of the LSS can be transformative in our study of Dark Energy.  There are three main reasons for this:

1. It allows direct measurement of the expansion history over the observed redshift range as explained in the previous section. Broad classes of dynamical Dark Energy exhibit ``tracking behaviour'' with respect to the dominant energy density at a given redshift \cite{Bull:2020cpe}, making measurements of the Dark Energy density during the transition into matter domination at $z\gtrsim 2$ particularly compelling. A combination of Baryon Acoustic Oscillations (BAO) and Redshift-Space Distortions (RSD) can directly obtain the Dark Energy density over the redshift range of observations \cite{Ferraro19}, thus severely constraining wide classes of models that mimic a cosmological constant at later times. In particular, Fig.\ \ref{fig:Omega_DE} shows both galaxy and 21cm experiments can constrain the fraction of Dark Energy to better than $2\%$ up to $z \approx 5$, fully covering the transition to matter domination.

2. It makes use of powerful degeneracy breaking: the parameter sensitivity varies considerably with redshift, and combining measurements over a wide redshift range can very effectively break degeneracies internally \cite{2020arXiv200708991E}.  This includes distinguishing the effects of dynamical Dark Energy and neutrino masses or other particles.

\begin{figure}[!h]
\centering
\includegraphics[width=0.9\linewidth]{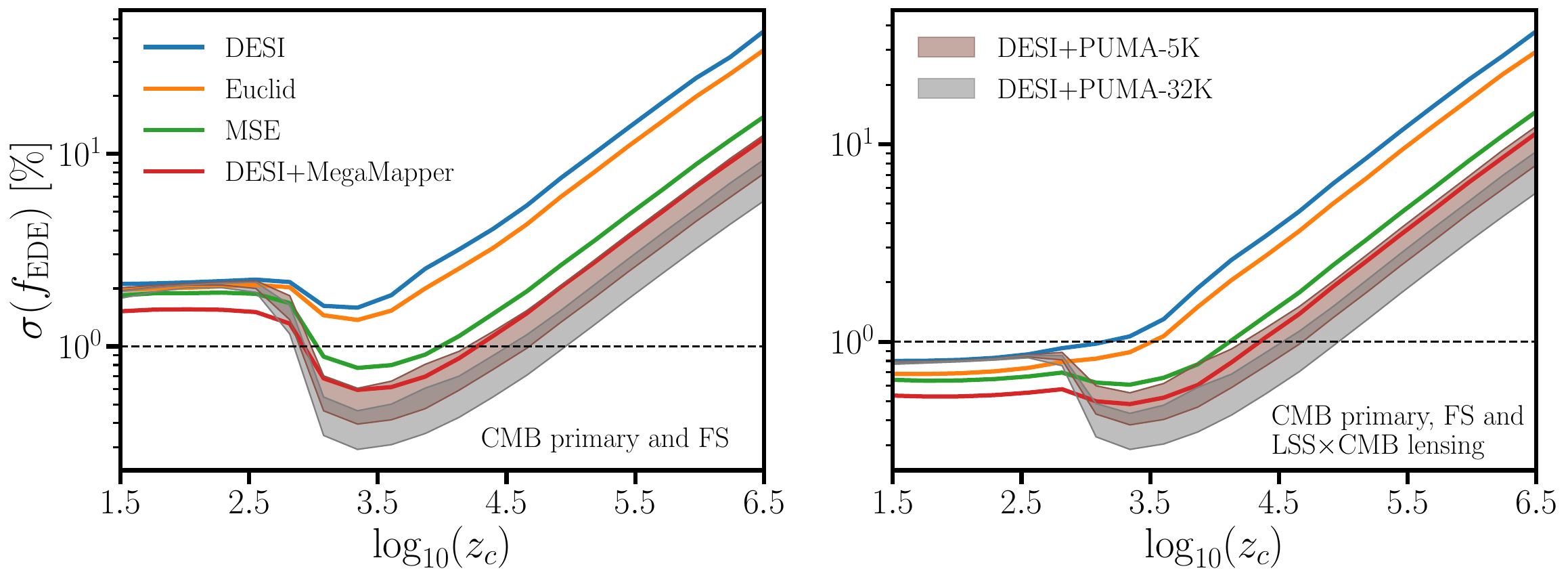}
\caption{Constraints on the maximum amplitude of early dark energy ($f_\text{EDE}$) as a function of the time at which EDE peaks $z_c$, assuming $\theta_i=2.83$. We include a Planck+SO prior on $\Lambda$CDM for all experiments. In the left panel we show constraints from full shape (FS) measurements only, while in the right panel we include a prior on $\Lambda$CDM and  nuisance parameters from SO lensing and cross-correlations with the respective galaxy surveys.
}
\label{fig:EDE_constraints}
\end{figure}

3. It is also an excellent indirect probe of early Dark Energy (EDE) to very high redshift: changes in the expansion rate at high $z$ temporarily alter the growth of structure and manifest themselves as features in the measured power spectrum \cite{Poulin:2018cxd, Hill20, Chen20b, Ivanov20}. By observing a very large volume, we can measure the shape of the power spectrum to an unprecedented accuracy (due to the reduced cosmic variance), and hence dramatically improve our sensitivity to EDE (and light relics, as probed by $N_{\rm eff}$ \cite{Baumann:2017gkg}). Quite excitingly, as shown in Fig.\ \ref{fig:EDE_constraints}, next-generation LSS surveys can constrain the fraction of EDE to be below 1\% all the way to $z \sim 10^5$, making them precision Dark Energy probes throughout most of cosmic history \cite{Sailer:2021}.

\section{Neutrinos and other extensions to the standard model}
\label{sec:extensions}

\begin{figure}[ht!]
\centering
\includegraphics[width=\linewidth]{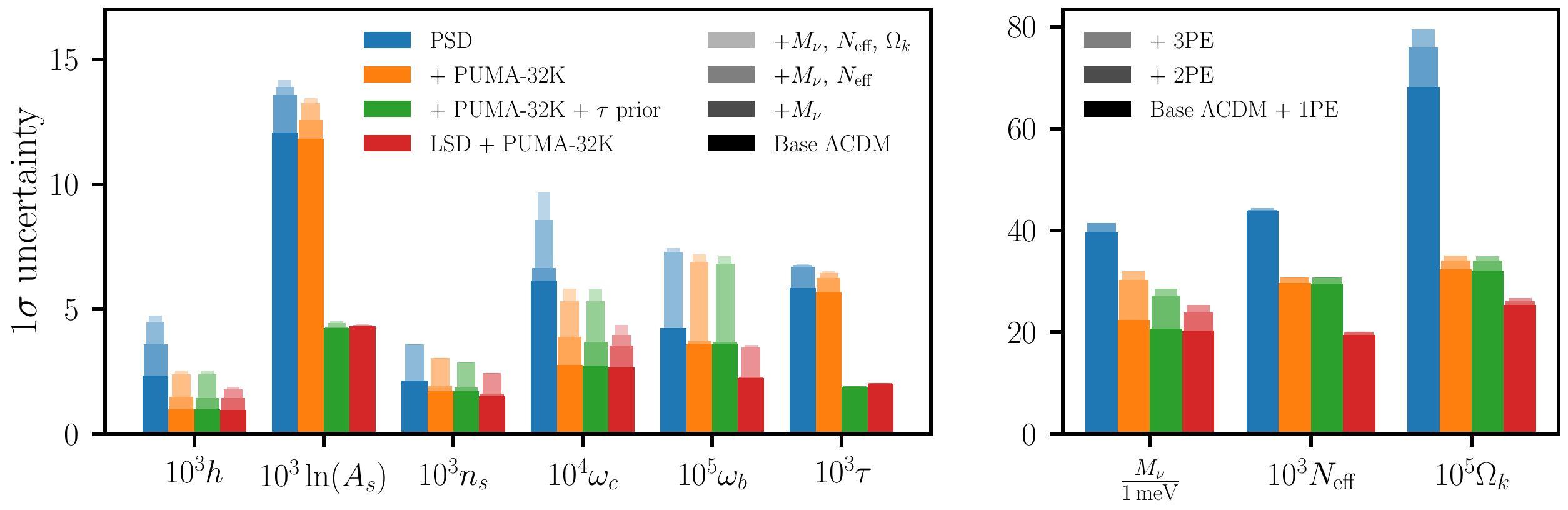}
\caption{Constraints on $\Lambda$CDM and extensions from four different combinations of experiments (starting with Planck+Simons+DESI=PSD or LiteBird+S4+DESI=LSD), using primary CMB and clustering data only. We include a $\sigma(\tau)=0.002$ prior for the green bars. \textit{Left}: The shading/width denotes which parameters are allowed to vary when fitting for $\Lambda$CDM (e.g.\ for Base $\Lambda$CDM+$M_\nu$ we keep $N_\text{eff}\,\text{and}\,\Omega_k$ fixed).
\textit{Right}: The shading denotes the number $N$ of Parameter Extensions ($N$PE) that are simultaneously allowed to vary in our fits, starting with the relevant parameter on the $x$-axis and continuing along the cycle $(M_\nu,N_\text{eff},\Omega_k)$; e.g.\ for $\Omega_k$, $2$PE is equivalent to $\Omega_k,\,M_\nu$.
}
\label{fig:extensions_to_base_model}
\end{figure}

\subsection{Massive Neutrinos}
\label{sec:neutrinos}

Massive neutrinos leave a number of imprints on the CMB and LSS, allowing their total mass to be measured through cosmological observations. The most important effect is a scale-dependent suppression of small scale power below the neutrino free-streaming scale. This suppression is proportional to the mass, and is redshift-dependent, allowing us to distinguish the effects of massive neutrinos from most models of dynamical Dark Energy. Probing a larger volume at high redshift allows a better determination of the large-scale power and hence of the amount of suppression, generally leading to future experiments constraining the overall sum of neutrino masses to $\sigma(M_\nu) \approx 20$ meV, or a $\approx 3\sigma$ ``detection'' in the case the minimum mass, normal hierarchy scenario (and more if the actual mass is larger). Figure \ref{fig:extensions_to_base_model} shows the forecasts in a number of scenarios.

\subsection{Light relativistic particles}

The presence of extra relativistic species at early times changes the radiation density (and hence affects the damping of the power spectrum on small scales), and also shifts the positions of the BAO peaks.  Therefore the high-precision measurements of the power spectrum by future high-redshift surveys is an ideal way to detect the presence of extra light particles.  Such observational constraints on the light relic density are expected to have broad implications for fundamental physics \cite{Alexander16,Green19}.  When parameterizing the number extra species by $N_{\rm eff}$ \cite{CMBS4}, the Standard Model of particle physics predicts $N_{\rm eff} = 3.046$ \cite{Mangano_2005}, and lower limits exist on the change $\Delta N_{\rm eff}$ for broad classes of particles: $\Delta N_{\rm eff} > $ 0.027 for a single scalar, 0.047 for a Weyl fermion, and 0.054 for vector boson, even if thermal decoupling happens earlier than the rest of the Standard Model\footnote{And a larger contribution if decoupling happens later than some of the Standard Model phase transitions.}. As we can see from Fig. \ref{fig:extensions_to_base_model}, future high-$z$ surveys can constrain $\sigma(N_{\rm eff})\approx 0.03$, and 0.02 when combined with proposed CMB experiments, reaching a sensitivity comparable to the smallest allowed value of $\Delta N_{\rm eff}$ for a single particle. While the constraints from LSS and a Stage-IV CMB experiment are comparable, we note that the amount of damping in the CMB due to $N_{\rm eff}$ is rather degenerate with the primordial Helium abundance $Y_p$, potentially degrading the constraints by up to a factor of 2 if tight external priors are not present \cite{Baumann:2017lmt, Baumann18}.  LSS measurements are not affected by $Y_p$ and the combination of LSS and CMB will provide the tightest and most robust constraints.

\subsection{Dark matter}

Although we do not provide specific forecasts for any given survey here, we would be remiss if we failed to mention that surveys of the type we are discussing would be able to provide exceptionally tight constraints on a range of postulated dark matter candidates \cite{Gluscevic19}.  Some of these constraints follow from similar considerations to the $N_{\rm eff}$ constraints described above (e.g.\ refs.~\cite{Green17,Green21}) while others provide more dramatic changes to the observables.  Examples of non-standard models that could be constrained by high-precision power spectrum measurements include those with dark matter-baryon scattering \cite{Dvorkin14}, dark matter interactions \cite{Leagourgues16,Pan18,Archidiacono19} or ultra-light axions \cite{Hlozek15} plus the very broad class of DM models that can be described with the effective theory of structure formation (ETHOS) framework \cite{ETHOS}.  More information about dark matter constraints from cosmic surveys can be found in a series of white papers submitted to the Snowmass Dark Matter: Cosmic Probes topical group \cite{DMsimsWP, DMhalosWP, DMPBHWP, DMextremeWP, DMfacilitiesWP}.

\section{Synergies with CMB and cross-correlations}
\label{sec:synergies}

While high redshift spectroscopic surveys provide compelling science reach on their own, it is also worth noting that the noise on CMB lensing maps from future wide-field experiments such as the Simons Observatory (SO) \cite{SimonsObs}, CMB-S4 \cite{CMBS4}, CMB-HD \cite{Sehgal:2019ewc} or PICO \cite{NASAPICO:2019thw}, will be reduced by more than one order of magnitude over existing measurements.  Unlike cosmic shear, CMB lensing can be measured to very high redshift, providing us with access to the matter field without the need to model bias.  Unfortunately CMB lensing alone mostly provides information that is projected along the line-of-sight (and with a broad redshift kernel).  The cross-correlation of CMB lensing with LSS in several redshift bins (CMB lensing tomography; \cite{Yu:2018tem, Yu:2021vce}), can break the degeneracy with galaxy bias inherent in LSS alone and provide measurement of the amplitude of perturbations as a function of redshift that cannot be obtained directly from the CMB.  This leads to tighter constraints on neutrino masses and Dark Energy, while mitigating some of the possible systematics \cite{Yu:2018tem,Wilson:2019brt,CMBS4}. Moreover, comparing the motion of non-relativistic matter through redshift-space distortions to the deflection of CMB photons will put some of the most informative bounds on theories of modified gravity \cite{Jain10}. At the same time, the cross-correlation of LSS with CMB lensing can potentially improve the robustness of constraints relying on the ultra-large scales, such as measurements of local non-Gaussianity \cite{Schmittfull:2017ffw}.

\begin{figure}[!h]
    \centering
    \includegraphics[width=\linewidth]{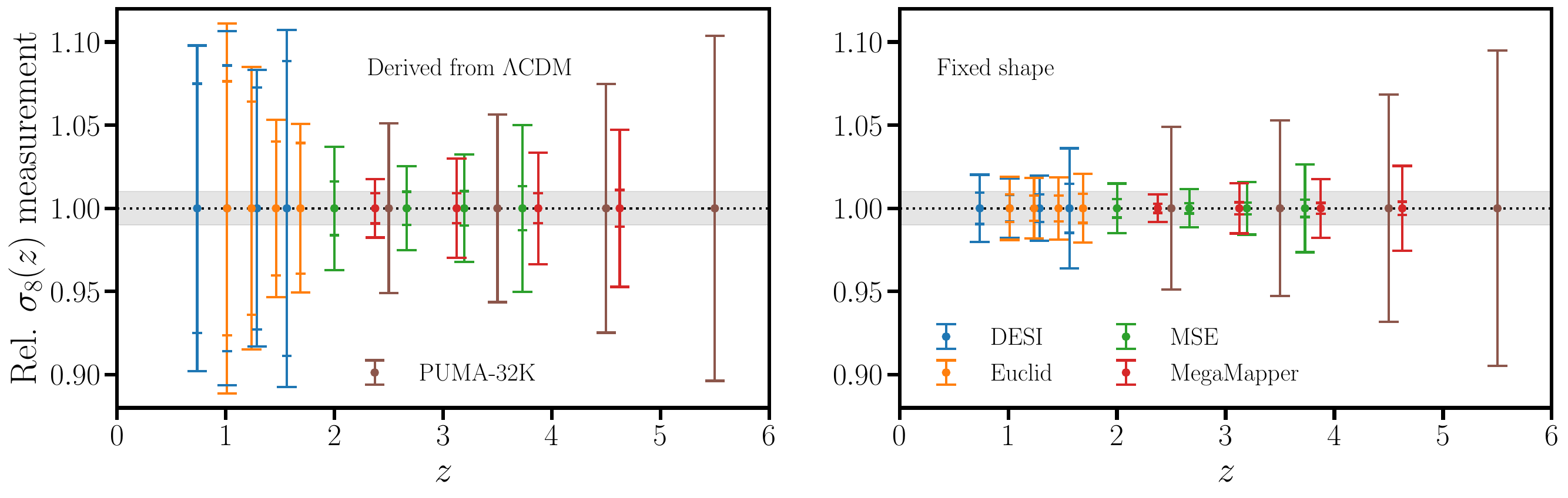}
    \caption{
    \textit{Left:} Errors on $\sigma_8(z)$ when derived from $\Lambda$CDM, assuming a $\sigma(\omega_b)=0.0005$ prior from BBN. \textit{Right:} Errors on $\sigma_8(z)$ from full shape data when calculated using the fixed shape procedure, without any external priors. The errorbars with smaller caps include cross-correlations with CMB-S4 lensing, but do not include any information from the convergence auto-spectrum. All 21-cm measurements assume optimistic foregrounds. The gray bands extend from $1\pm0.01$.
    }
\label{fig:fs8}
\end{figure}

In addition to CMB lensing, other secondary anisotropies\footnote{That is, fluctuations caused by the interaction of CMB photons with matter along the line of sight.} of the CMB correlate CMB and LSS maps, revealing the properties of the gas in the cosmic web (through the thermal and kinematic Sunyaev-Zel'dovich effects, tSZ and kSZ \cite{1972CoASP...4..173S, SZ80, Battaglia:2017neq}), and providing a tool to measure large-scale halo velocities through the kSZ \cite{Smith:2018bpn} and ``moving lens'' \cite{1983Natur.302..315B, Hotinli:2018yyc} effects. These velocity measurements can provide access to the very largest scales, often the most affected by primordial physics. In combination with LSS, they can be used to reduce cosmic variance, potentially providing an independent measurement of local primordial non-Gaussianity with $\sigma(f_{\rm NL}^{\rm loc}) \lesssim 1$ \cite{Munchmeyer:2018eey}.  To make the most out of these opportunities, the arcminute or sub-arcminute resolution of future ground-based CMB experiments will be essential.

\section{Figure of Merit}
\label{sec:FOM}

Essentially all constraints derived in this white paper are connected to extracting information from linear (or weakly non-linear) modes. These are modes for which coupling with other modes can be assumed to be sub-dominant and which have evolved largely unchanged (except for the linear growth) from the early stages of the universe. Non-linear coupling mixes information from different modes and typically erases any pristine features that were present. Therefore it makes sense to establish a Primordial Physics Figure of Merit that counts those modes. The mathematical details are discussed in appendix \ref{app:FOM}, but the effective number of modes are in essence computed by calculating the total number of modes accessible to the survey and weighting them by the wave-number dependent kernel that measures how much information has been lost due to i) non-linear evolution and ii) the noise properties of the survey (dominated by Poisson noise for galaxy surveys and receiver thermal noise for intensity mapping surveys). The resulting quantity converges to a finite number without an externally imposed scale cut, e.g.\ $k_{\rm max}$. It is therefore a quantity that is survey dependent, but has otherwise no tunable parameters. 

Some of the physics discussed here, such as sensitivity to the features in the primordial power spectrum, the constraints on early dark energy and constraints on primordial non-Gaussianity (with some important caveats) scale almost linearly with this quantity (in the sense that errors scale proportionally to the FoM$^{-\sfrac{1}{2}}$) and allow us to compare survey proposals directly. Other science, such as BAO and neutrino mass constraints still scale approximately this way \cite{Sailer:2021}. We therefore think it is a good candidate to replace the Dark Energy Figure Of Merit, which has outlived its usefulness in the 2020s.

As with any Figure of Merit, there are limitations in compressing the information into a single number. Some are known unknowns. For example, we know that 21\,cm has to filter the slowly varying line-of-sight modes. This limits its usefulness in cross-correlations with 2D tracers (most importantly CMB secondaries, but also galaxy shear maps -- but see \cite{Modi19}) and limits its reach alone for local non-Gaussianity in a way that is not captured by simply omitting those modes from the count.  There are also unknown unknowns. We know that galaxy survey spectroscopy is a well tested method with a long track record of delivering the forecasted science. Intensity mapping techniques are considerably less technically advanced and, while promising, their high FoM has be judged against the lower, but more robust, FoM forecasted for spectroscopic galaxy surveys. 

In Table \ref{tab:experiments} we enumerate some of the current experiments relevant for this white-paper together with their FoM. Taking the emission line galaxy (ELG) sample from DESI as a reference, we see that all future experiments (with the exception of MSE) allow for at least an order of magnitude improvement in its value. As expected, we find that the more aggressive designs yields bigger improvements: if SpecTel concentrated entirely on large-$N$ science it would be over a factor of two better than MegaMapper. PUMA in its ultimate configuration is two and a half order of magnitude better than DESI and a factor of 40 better than MegaMapper. This is of course contingent on 21\,cm delivering results that are systematically clean commensurate with its statistical sensitivity, which is not the case for the current generation of 21\,cm experiments \cite{CHIME22}.

\section{Conclusions}

Measurements of the inhomogeneous Universe with the CMB and LSS currently provide our tightest constraints on inflation, neutrino masses, light relic particles, dark energy and dark matter.  \textbf{New experimental techniques allow us to dramatically extend the redshift frontier. By probing large-scale structure up to $z\approx 5.5$ we will more than  quadruple the observed volume, dramatically increasing the number of observed modes and thus enable large improvements on our understanding of the primordial Universe.}  These observations will build upon community investments in optical and sub-mm facilities about to come on line and will allow us to probe the metric, particle content and both epochs of accelerated expansion with high precision.

The high redshift Universe, where we can accurately measure long-wavelength modes, offers a theoretically robust route to the constraints discussed above.  Resting upon perturbative models that are built around principles familiar from high-energy particle physics, our ability to forecast and optimize these large-N surveys is well developed and theoretically secure.  \textbf{Such surveys aim at the location where the inference is the cleanest and the noise lowest, maximizing the discovery potential for beyond standard model physics in a theory-agnostic manner.}  This is also the regime where our forecasts are most reliable, allowing more in-depth survey optimization.

\textbf{Precise studies of large-scale structure contribute to all of the topics in the cosmic frontier, including constraints on the expansion history and curvature, multiple probes of inflation, limits on dark energy across the whole history of the Universe, the properties of dark matter and its interactions, on light relics and neutrino mass and provide constraints on modified gravity.}
Future experiments will measure the expansion history to sub-percent precision all the way to $z\simeq 5$, fully spanning the dark energy dominated era and extending well into matter domination.  By using the shape of the power spectrum future surveys could constrain the fraction of early dark energy to be $<1\%$ all the way to $z\simeq 10^5$.
The combination of large-scale structure and CMB experiments should allow detection of the minimum mass, normal neutrino hierarchy at $\approx 3\sigma$.
They constrain the light relic particle density to a level comparable to the smallest allowed value for any single particle (regardless of spin).
Primordial non-Gaussianity of the local type is forecast to be measured with $\sigma(f_{\rm NL}^{\rm loc})<1$, allowing separation of single- and multi-field inflationary models, crossing an important theoretical threshold and reaching beyond what is achievable with the CMB even in principle.

\textbf{The combination of spectroscopic measurements of large-scale structure in the high-redshift Universe and deep CMB observations is particularly powerful.}  Forecasts suggest the combination leads to tighter constraints on neutrinos, light relics, dark energy, inflation and modified gravity while mitigating some possible systematics.

\section{Acknowledgements}
We thank Kyle Dawson, Alex Drlica-Wagner, Mustapha Ishak, Kirit Karkare, Juna Kollmeier, Azadeh Moradinezhad Dizgah, David Schlegel, Neelima Sehgal and the members of the Cosmic Frontier 4 group for very helpful discussions and comments on the draft. 

\appendix

\clearpage

\section{``Primordial Physics'' figure of merit}
\label{app:FOM}

Motivated by the scaling of the Fisher matrix for the reconstructing the primordial power spectrum, we follow \cite{Sailer:2021} and define the effective number of modes correlated with the initial conditions as follows:
While ``linear'' modes at low $k$ are 100\% correlated with the initial conditions because linear evolution preserves the phase information, non-linear evolution and measurement noise create decorrelation on smaller scales, leading to a loss of information. We model this decorrelation with a Gaussian propagator given by:
\begin{equation}
    G(k,\mu) \equiv\frac{\langle\delta_F\delta_L\rangle}{\langle\delta_L\delta_L\rangle}
    \simeq \exp\left[-\frac{1}{2}\left(k_\perp^2+k_\parallel^2\{1+f\}^2 \right)\Sigma^2\right],
\end{equation}
where $\Sigma$ is set by the rms displacement within the Zel'dovich approximation and $f$ is the linear growth factor.

We can split the nonlinear overdensities $\delta_F(\bm{k}) = G(\bm{k})\delta_L(\bm{k}) + d(\bm{k})$ into two pieces: one which is correlated with the linear overdensities, and one which isn't $\langle\delta_L(\bm{k}) d(\bm{k}')\rangle=0$. 
In the sample variance limit, the relative uncertainty of the linear power spectrum's amplitude $A$ is given by $\sqrt{2/N_{\rm modes}}$. We can calculate this uncertainty from the Fisher matrix:
\begin{equation}
    F_{ii}
    =
    \frac{f_{\rm sky}}{2}
    \int_{z_{\rm min}}^{z_{\rm max}} dz
    \frac{dV}{dz}
    \int_{k-{\rm wedge}(z)}^\infty
    \frac{d^3 \bm{k}}{(2\pi)^3}
    \left(
    \frac{
    \partial_{p_i} P_F(\bm{k},z)}{
    P_F(\bm{k},z) + N(z)}
    \right)^2
    ,
\end{equation}
where $k-{\rm wedge}(z)$ defines the lower limit of integration. The uncertainty on the amplitude is $\delta A/A = \delta P_L(\bm{k})/P_L(\bm{k}) = 1/\sqrt{F_{AA}}$, so that $N_{\rm modes}=2F_{AA}$. This defines the figure of merit $N_{\rm modes}$. A short calculation gives:
\begin{equation}
    N_{\rm modes}
    =
    f_{\rm sky}
    \int_{z_{\rm min}}^{z_{\rm max}} dz
    \frac{dV}{dz}
    \int_{k-{\rm wedge}(z)}^\infty
    \frac{d^3 \bm{k}}{(2\pi)^3}
    \left(
    \frac{
    G^2(\bm{k},z)P_L(\bm{k},z)}{
    P_F(\bm{k},z) + N(z)}
    \right)^2.
\end{equation}
To define a Primordial Physics Figure of Merit (Primordial FoM) of order unity, in this paper we take:
\begin{equation}
    {\rm Primordial \ FoM} \equiv 10^{-6} N_{\rm modes}
\end{equation}
Note that since the definition above does not include the sensitivity of the power spectrum to the parameters of interest, it is a useful quantity to compare different ``similar'' experiments on their sensitivity to primordial physics. However, for example, it should not be used to compare sensitivity to ``low-redshift'' parameters such as Dark Energy or neutrinos between CMB and LSS experiments. 

When numerically evaluating $N_{\rm modes}$, Ref. \cite{Sailer:2021} finds that high-$z$ LSS experiments can surpass the CMB, but for this to happen, observations at $z>3.5$ are necessary.

\clearpage
\bibliographystyle{unsrturltrunc6.bst}
\bibliography{main}

\end{document}